\documentclass[twocolumn,showpacs,preprintnumbers,amsmath,amssymb]{revtex4}
\begin{document}
\draft{}
\bibliographystyle{try}

\topmargin 0.1cm

 \newcounter{univ_counter}
 \setcounter{univ_counter} {0}

\addtocounter{univ_counter} {1} 
\edef\INFNGE{$^{\arabic{univ_counter}}$ } 

\addtocounter{univ_counter} {1} 
\edef\JLAB{$^{\arabic{univ_counter}}$ } 

\addtocounter{univ_counter} {1} 
\edef\MSU{$^{\arabic{univ_counter}}$ } 

\addtocounter{univ_counter} {1} 
\edef\ASU{$^{\arabic{univ_counter}}$ } 

\addtocounter{univ_counter} {1} 
\edef\UCLA{$^{\arabic{univ_counter}}$ } 

\addtocounter{univ_counter} {1} 
\edef\CMU{$^{\arabic{univ_counter}}$ } 

\addtocounter{univ_counter} {1} 
\edef\CUA{$^{\arabic{univ_counter}}$ } 

\addtocounter{univ_counter} {1} 
\edef\SACLAY{$^{\arabic{univ_counter}}$ } 

\addtocounter{univ_counter} {1} 
\edef\CNU{$^{\arabic{univ_counter}}$ } 

\addtocounter{univ_counter} {1} 
\edef\UCONN{$^{\arabic{univ_counter}}$ } 

\addtocounter{univ_counter} {1} 
\edef\DUKE{$^{\arabic{univ_counter}}$ } 

\addtocounter{univ_counter} {1} 
\edef\EDINBURGH{$^{\arabic{univ_counter}}$ } 

\addtocounter{univ_counter} {1} 
\edef\FIU{$^{\arabic{univ_counter}}$ } 

\addtocounter{univ_counter} {1} 
\edef\FSU{$^{\arabic{univ_counter}}$ } 

\addtocounter{univ_counter} {1} 
\edef\GWU{$^{\arabic{univ_counter}}$ } 

\addtocounter{univ_counter} {1} 
\edef\GLASGOW{$^{\arabic{univ_counter}}$ } 

\addtocounter{univ_counter} {1} 
\edef\INFNFR{$^{\arabic{univ_counter}}$ } 

\addtocounter{univ_counter} {1} 
\edef\ORSAY{$^{\arabic{univ_counter}}$ } 

\addtocounter{univ_counter} {1} 
\edef\ITEP{$^{\arabic{univ_counter}}$ } 

\addtocounter{univ_counter} {1} 
\edef\JMU{$^{\arabic{univ_counter}}$ } 

\addtocounter{univ_counter} {1} 
\edef\KYUNGPOOK{$^{\arabic{univ_counter}}$ } 

\addtocounter{univ_counter} {1} 
\edef\MIT{$^{\arabic{univ_counter}}$ } 

\addtocounter{univ_counter} {1} 
\edef\UMASS{$^{\arabic{univ_counter}}$ } 

\addtocounter{univ_counter} {1} 
\edef\UNH{$^{\arabic{univ_counter}}$ } 

\addtocounter{univ_counter} {1} 
\edef\NSU{$^{\arabic{univ_counter}}$ } 

\addtocounter{univ_counter} {1} 
\edef\OHIOU{$^{\arabic{univ_counter}}$ } 

\addtocounter{univ_counter} {1} 
\edef\ODU{$^{\arabic{univ_counter}}$ } 

\addtocounter{univ_counter} {1} 
\edef\PITT{$^{\arabic{univ_counter}}$ } 

\addtocounter{univ_counter} {1} 
\edef\ROMA{$^{\arabic{univ_counter}}$ } 

\addtocounter{univ_counter} {1} 
\edef\RPI{$^{\arabic{univ_counter}}$ } 

\addtocounter{univ_counter} {1} 
\edef\RICE{$^{\arabic{univ_counter}}$ } 

\addtocounter{univ_counter} {1} 
\edef\URICH{$^{\arabic{univ_counter}}$ } 

\addtocounter{univ_counter} {1} 
\edef\SCAROLINA{$^{\arabic{univ_counter}}$ } 

\addtocounter{univ_counter} {1} 
\edef\UTEP{$^{\arabic{univ_counter}}$ } 

\addtocounter{univ_counter} {1} 
\edef\VT{$^{\arabic{univ_counter}}$ } 

\addtocounter{univ_counter} {1} 
\edef\VIRGINIA{$^{\arabic{univ_counter}}$ } 

\addtocounter{univ_counter} {1} 
\edef\WM{$^{\arabic{univ_counter}}$ } 

\addtocounter{univ_counter} {1} 
\edef\YEREVAN{$^{\arabic{univ_counter}}$ } 

\title{{\large Measurement of $ep \rightarrow e^{'}p\pi^{+}\pi^{-}$ 
and baryon resonance analysis}}
 \author{ 
M.~Ripani,\INFNGE\
V.D.~Burkert,\JLAB\
V.~Mokeev,\MSU\
M.~Battaglieri,\INFNGE\
R.~De~Vita,\INFNGE\
E.~Golovach,\MSU\
M.~Taiuti,\INFNGE\
G.~Adams,\RPI\
E.~Anciant,\SACLAY\
M.~Anghinolfi,\INFNGE
% ...
} 

\address{\INFNGE INFN, Sezione di Genova, 16146 Genova, Italy}
\address{\JLAB Thomas Jefferson National Accelerator Facility, Newport News, Virginia 23606}
\address{\MSU Moscow State University, 119899 Moscow, Russia}
\address{\ASU Arizona State University, Tempe, Arizona 85287}
\address{\UCLA University of California at Los Angeles, Los Angeles, California  90095}
\address{\CMU Carnegie Mellon University, Pittsburgh, Pennsylvania 15213}
\address{\CUA Catholic University of America, Washington, D.C. 20064}
\address{\SACLAY CEA-Saclay, Service de Physique Nucl\'eaire, F91191 Gif-sur-Yvette, Cedex, France}
\address{\CNU Christopher Newport University, Newport News, Virginia 23606}
\address{\UCONN University of Connecticut, Storrs, Connecticut 06269}
\address{\DUKE Duke University, Durham, North Carolina 27708}
\address{\EDINBURGH Edinburgh University, Edinburgh EH9 3JZ, United Kingdom}
\address{\FIU Florida International University, Miami, Florida 33199}
\address{\FSU Florida State University, Tallahasee, Florida 32306}
\address{\GWU The George Washington University, Washington, DC 20052}
\address{\GLASGOW University of Glasgow, Glasgow G12 8QQ, United Kingdom}
\address{\INFNFR INFN, Laboratori Nazionali di Frascati, PO 13, 00044 Frascati, Italy}
\address{\ORSAY Institut de Physique Nucleaire ORSAY, IN2P3 BP 1, 91406 Orsay, France}
\address{\ITEP Institute of Theoretical and Experimental Physics, Moscow, 117259, Russia}
\address{\JMU James Madison University, Harrisonburg, Virginia 22807}
\address{\KYUNGPOOK Kyungpook National University, Daegu 702-701, South Korea}
\address{\MIT Massachusetts Institute of Technology, Cambridge, Massachusetts  02139}
\address{\UMASS University of Massachusetts, Amherst, Massachusetts  01003}
\address{\UNH University of New Hampshire, Durham, New Hampshire 03824}
\address{\NSU Norfolk State University, Norfolk, Virginia 23504}
\address{\OHIOU Ohio University, Athens, Ohio 45701}
\address{\ODU Old Dominion University, Norfolk, Virginia 23529}
\address{\PITT University of Pittsburgh, Pittsburgh, Pennsylvania 15260}
\address{\ROMA Universita' di ROMA III, 00146 Roma, Italy}
\address{\RPI Rensselaer Polytechnic Institute, Troy, New York 12180}
\address{\RICE Rice University, Houston, Texas 77005}
\address{\URICH University of Richmond, Richmond, Virginia 23173}
\address{\SCAROLINA University of South Carolina, Columbia, South Carolina 29208}
\address{\UTEP University of Texas at El Paso, El Paso, Texas 79968}
\address{\VT Virginia Polytechnic Institute and State University, Blacksburg, Virginia   24061}
\address{\VIRGINIA University of Virginia, Charlottesville, Virginia 22901}
\address{\WM College of William and Mary, Williamsburg, Virginia 23187}
\address{\YEREVAN Yerevan Physics Institute, 375036 Yerevan, Armenia}
%
 
%
%

\date{\today}

\begin{abstract}
The cross section for the reaction $ e p \rightarrow e^{\prime} p \pi^{+} \pi^{-}$ 
was measured in the resonance region for 1.4$<$W$<$2.1 GeV and
0.5$<Q^{2}<$1.5 GeV$^{2}$/c$^{2}$ using the CLAS detector at Jefferson Laboratory.
The data shows resonant 
structures not visible in previous experiments.
The comparison of our 
data to a phenomenological prediction using available information
on $N^{*}$ and $\Delta$ states shows an evident discrepancy.
A better description of the data is obtained either by a sizeable change of the
properties of the $P_{13}$(1720) resonance or by introducing a new
baryon state, not reported in published analyses. 
\end{abstract}

\pacs{13.60.Le, 13.40.Gp, 14.20.Gk}

\maketitle

Electromagnetic excitation of nucleon resonances is sensitive to the 
spin and spatial structure of the transition, which in turn is
connected to fundamental properties of baryon structure,
like spin-flavor symmetries, confinement, and 
effective degrees of freedom.
In the mass region above 1.6 GeV, many overlapping baryon states
are present, and some of them are not well known;  
measurement of the transition form factors of these states 
is important for our understanding of the internal dynamics
of baryons.
Many of these high-mass excited states 
tend to decouple from the single-meson 
channels and to decay predominantly 
into multi-pion channels, such as $\Delta \pi$ or $N \rho$, leading 
to $N \pi \pi$  final states \cite{Pdg96}. 
Moreover, quark models
with approximate (or ``broken'') SU(6)$\otimes$O(3) symmetry \cite{Kon80,Gia90} 
predict more states than have been found experimentally; 
QCD mixing effects could decouple these unobserved states 
from the pion-nucleon channel \cite{Kon80} while
strongly coupling them to two-pion channels \cite{Kon80,Cap94,Sta93}. 
These states would therefore not be observable in reactions with
$\pi N$ in the initial or final state.
Other models, with different symmetry properties and a reduced
number of degrees of freedom, as e.g. in ref. \cite{Kir97}, 
predict fewer states. 
Experimental searches for at least some of the
   ``missing'' states predicted by the symmetric quark models, which are
   not predicted by models using alternative symmetries, are crucial in
   discriminating between these models.
Electromagnetic amplitudes for some missing states are predicted
to be sizeable \cite{Kon80} as well.
Therefore, exclusive double-pion electroproduction 
is a fundamental tool in measuring poorly known states and 
possibly observing new ones. 

In this paper we report a measurement 
of the $e p \rightarrow e^{\prime} p \pi^{+} \pi^{-}$ reaction 
studied with the CEBAF Large Acceptance Spectrometer (CLAS) 
at Jefferson Lab. Beam currents of 
a few nA were delivered to Hall B on a liquid-hydrogen target, 
corresponding to luminosities up to $4 \times 10^{33}$~cm$^{-2}$s$^{-1}$. 
Data were
taken in 1999 for about two months at beam energies of 2.6 and 4.2 GeV.
Important features
of the CLAS \cite{Bro99} are its large kinematic coverage for 
multi-charged-particle final states
and its good momentum resolution ($\Delta p/p \sim$1\%). Using an
inclusive electron trigger based on a coincidence between the forward
electromagnetic shower calorimeter and the gas Cerenkov detector,
many exclusive hadronic final states
were measured simultaneously. Scattered electrons were identified through
cuts on the calorimeter energy loss and the Cerenkov photo-electron distribution.
Different channels were separated through
particle identification using time-of-flight information and other 
kinematic cuts.
We used the missing-mass technique, 
requiring detection in CLAS of at least $e p \pi^{+}$.
The good resolution allowed selection of the exclusive
   final state, $ep \pi^{+} \pi^{-}$.  After applying all cuts, our data sample
   included about $2 \times 10^{5}$ two-pion events. 
%
%

The range of invariant hadronic center-of-mass (CM) 
energy $W$ (in 25 MeV bins) was
1.4-1.9~GeV for the first two bins in the invariant momentum
transfer $Q^{2}$, 0.5-0.8~(GeV/c)$^2$ and 0.8-1.1~(GeV/c)$^2$, and 
1.4-2.1~GeV for the highest $Q^{2}$ bin, 1.1-1.5~(GeV/c)$^2$. 
Data were corrected for acceptance, reconstruction efficiency, 
radiative effects, and 
empty target counts. They were further binned in the following 
set of hadronic CM variables: 
invariant mass of the $p \pi^{+}$ pair (10 bins), invariant mass 
of the $\pi^{+} \pi^{-}$ pair (10 bins), 
$\pi^{-}$ polar angle $\theta$ (10 bins), azimuthal angle $\phi$ (5 bins), 
and rotation freedom $\psi$ of the $p \pi^{+}$ pair with respect to
the hadronic plane (5 bins).
The full differential cross section is
of the form: 
\begin{eqnarray} 
\frac{d\sigma}{dWdQ^{2}dM_{p \pi^{+}}dM_{\pi^{+}\pi^{-}}d\cos\theta_{\pi^{-}}d\phi_{\pi^{-}}d\psi_{p \pi^{+}}} = \nonumber \\
\Gamma_{v}\frac{d\sigma_{v}}{dM_{p \pi^{+}}dM_{\pi^{+}\pi^{-}}d\cos\theta_{\pi^{-}}d\phi_{\pi^{-}}d\psi_{p \pi^{+}}} =
\Gamma_{v}\frac{d\sigma_{v}}{d\tau}
\end{eqnarray} 
\begin{equation} 
\Gamma_{v} =
\frac{\alpha}{4 \pi}\frac{1}{E^{2}M_{p}^{2}}
\frac{W(W^{2}-M_{p}^{2})}{(1-\epsilon)Q^{2}}
\end{equation}
\noindent where $\Gamma_{v}$ is the virtual photon flux,
$\frac{d\sigma_{v}}{d\tau}$ is the virtual photon cross section,
$\alpha$ is the fine structure constant,
$E$ is the electron beam energy,
$M_{p}$ is the proton mass, and $\epsilon$ is the virtual photon
transverse polarization \cite{Ama89}.
\begin{figure}[h]
\vspace{5.cm}                                                                  
\includegraphics{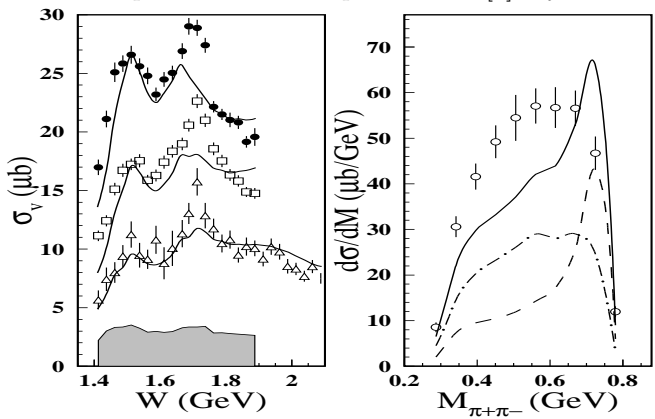}
\caption[]{Left: Total cross section for 
$\gamma_{v} p \rightarrow p \pi^{+} \pi^{-}$ as a function of $W$.
Data from CLAS are shown at $Q^{2}$=0.5-0.8~(GeV/c)$^2$ (full points),
$Q^{2}$=0.8-1.1~(GeV/c)$^2$ (open squares), and
$Q^{2}$=1.1-1.5~(GeV/c)$^2$ (open triangles). 
Error bars are statistical only, while the bottom band shows the sytematic
error for the lowest $Q^{2}$ bin.
The curves represent our step (A) reference calculations.
Right: $\frac{d\sigma_{v}}{dM_{\pi^{+}\pi^{-}}}$ from CLAS
at $Q^{2}$=0.8-1.1~(GeV/c)$^2$ and $W$=1.7-1.725 GeV
(statistical error bars only). 
The curves represent our step (A) reference calculations,
extrapolated to the edge points.
The dashed line includes all resonances,
the dot-dashed line includes only the non-resonant part, and
the solid line is the full calculation.
}
\label{fig:Xsec_allq2_data_and_nominal_and_pipi}
\end{figure}
Systematic uncertainties were estimated as a function of $W$ and $Q^{2}$.
The main sources were acceptance modeling, finite integration steps,
and modeling of the radiative corrections, each one being at the 3 to 10\% level. 
Each of the various cuts applied (fiducial, missing mass, etc.) contributed 2 to 5\%.
In Fig.~\ref{fig:Xsec_allq2_data_and_nominal_and_pipi} (left)
we report the total virtual photon cross section as a function of
$W$ for all $Q^{2}$ intervals analyzed. 
The CLAS data points clearly exhibit structures, 
not visible in previous data \cite{Eck73}
due to limited statistical accuracy.

Since existing theoretical models \cite{Ose00} are limited
to $W<$1.6 GeV,
we have employed a phenomenological calculation \cite{Mok01}
for a first
   interpretation of the data.  This model describes the reaction 
   $\gamma_{v} p \rightarrow p \pi^{+} \pi^{-} $ in the
kinematic range of interest as a sum of amplitudes for
$\gamma_{v} p \rightarrow \Delta \pi \rightarrow p \pi^{+} \pi^{-} $ and 
$\gamma_{v} p \rightarrow \rho^{0} p \rightarrow p \pi^{+} \pi^{-} $,
while all other possible mechanisms are
parameterized as phase space. 
A detailed treatment was developed 
for the non-resonant
contributions to $\Delta \pi $, while for $\rho p$
production they were described through a simple diffractive ansatz.
For the resonant part, a total of 12 states, classified as
4$^{*}$ \cite{Pdg96}, with sizeable $\Delta \pi$ and/or $\rho p$ 
decays, were included based on a Breit-Wigner ansatz.
A few model parameters in non-resonant production
were fitted to CLAS data at high $W$, where the non-resonant
part creates a forward
peaking in the angular distributions, and kept fixed 
in the subsequent analysis. The phase between resonant 
and non-resonant $\Delta \pi $ mechanisms 
was fitted to the CLAS data as well.
To simplify the fits, we reduced eqn. (1) 
to three single-differential cross sections,
$\frac{d \sigma}{dM_{p \pi^+}}$,
   $\frac{d \sigma}{dM_{\pi^+ \pi^-}}$, and $\frac{d \sigma}{d \cos\theta_{\pi^-}}$,
   by integrating over the other hadronic variables.  
   These three 1-D distributions were 
   then fitted simultaneously.  
   Here $\frac{d \sigma}{dM_{p \pi^+}}$ and
   $\frac{d \sigma}{d \cos\theta_{\pi^-}}$ are both connected with the
   dominant $\Delta^{++} \pi^-$ production reaction, while
   $\frac{d \sigma}{dM_{\pi^+ \pi^-}}$ is connected with $p \rho^0$
   production.
For each $W$ and $Q^2$ bin, a total of 26
   data points from the three single-differential cross sections
   were used in our fits. The two edge points in both the $p \pi^{+}$ and
   $\pi^{+} \pi^{-}$ mass distributions were excluded as the model did not
   take into account the kinematic smearing in the $M_{\pi^{+} \pi^{-}}$ versus
   $M_{p \pi^{+}}$ plot caused by the $W$ bin width.

The data analysis was performed in the following steps:
(A) We produced reference curves using the available 
information on the
   $N^*$ and $\Delta$ resonances in 1.2-2 GeV mass range.
Discrepancies between the CLAS data and our calculation were observed,
which led to the subsequent steps B and C.
(B) Data around $W$=1.7 GeV were fitted using the 
   known resonances in the PDG but allowing the resonance parameters 
   to vary in a number of ways.  The best fit, corresponding to a
   prominent $P_{13}$ partial wave, could be
attributed to the PDG $P_{13}$(1720) resonance, but with parameters significantly modified
   from the PDG values.
(C) As an alternative description, we introduced a new baryon state around
1.7~GeV.
In what follows we describe each of the steps above in more detail.

Step (A) - To produce our reference curves,
   the $Q^2$ evolution of the $A_{1/2}$ and $A_{3/2}$ electromagnetic
   couplings for the states was taken either from parameterizations
   of existing data \cite{Bur94}, or from 
Single Quark Transition Model (SQTM) fits \cite{Bur94} where no data was available.
For the $P_{11}$(1440) (Roper), given the scarce available data,
the amplitude $A_{1/2}$ was taken from a 
Non-Relativistic Quark Model (NRQM) \cite{Clo90}.
Partial $LS$ decay widths were taken from
a previous analysis of hadronic data \cite{Man92} and renormalized
to the total widths from Ref. \cite{Pdg96}.
Results for step (A) are reported in Fig.~\ref{fig:Xsec_allq2_data_and_nominal_and_pipi} (left).
The strength of the $D_{13}$(1520) resonance at 1.5 GeV and the underlying
continuum are well reproduced, except for the region on the low-mass side
of the peak. 
However, a strong discrepancy is evident at $W$ around 1.7 GeV.
Moreover, at this energy the reference curve
exhibits a strong peak in the $\pi^+ \pi^-$
   invariant mass 
(Fig.~\ref{fig:Xsec_allq2_data_and_nominal_and_pipi}, right), 
connected to sizeable
$\rho$ meson production. This contribution was traced back to the 70-85\% 
branching ratio of the $P_{13}$(1720) into this channel \cite{Pdg96,Man92,Dyt00}. 

Step (B) - Starting from the above mentioned reference values, 
the parameters of various states were varied in order
to fit the CLAS data. In this discussion, we restrict
ourselves to the discrepancy around 1.7 GeV and the few
resonant states relevant in this energy region.
All fit $\chi^2/\nu$ values were calculated from the 8 $W$ bins 
between 1.64 and 1.81 GeV and from the 3 $Q^{2}$ bins (624 data points). 
The number of free parameters ranged from 11 to 32,
depending on the fit,
corresponding to $\nu$=613 to 592 degrees of freedom.  
Assuming the resonance properties given by the PDG,
the bump at
   about $W$=1.7 GeV cannot be due to the
$D_{15}$(1675), $F_{15}$(1680), or $D_{33}$(1700) states; 
the first because its well known position cannot match the peak;
the second because of its well known position and 
photocouplings \cite{Bur01};
the third due to its large width ($\sim$300 MeV).
The remaining possibilities from the PDG 
are the $D_{13}$(1700), the $P_{13}$(1720), and the $P_{11}$(1710)
(the latter not present in step (A)), 
for which no data on $A_{1/2}$ or $A_{3/2}$ at $Q^{2}>0$ 
are available \cite{Bur01}. 
If no configuration mixing 
occurs, the $D_{13}$(1700) cannot be excited in the SQTM,
while the SQTM prediction for the $P_{13}$(1720) 
relies on ad hoc assumptions.
According to the literature \cite{Pdg96,Man92,Dyt00}, 
hadronic couplings of the $D_{13}$(1700) and the total width
of the $P_{11}$(1710)
are poorly known, while the $P_{13}$(1720) hadronic
parameters should be better established.
Several other partial waves were investigated
in step (C). 
%
\begin{figure}[h]
\vspace{6.0cm}                                                                  
\includegraphics{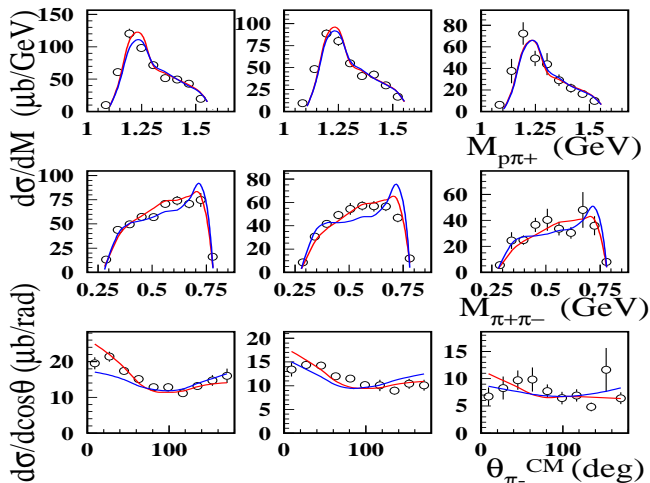}
\caption[]{$\frac{d\sigma_{v}}{dM_{p \pi^{+}}}$, 
$\frac{d\sigma_{v}}{dM_{\pi^{+}\pi^{-}}}$, 
and $\frac{d\sigma_{v}}{dcos\theta_{\pi^{-}}}$
from CLAS (from top to bottom) at $W$=1.7-1.725~GeV
and for the three mentioned $Q^{2}$ intervals (left to right).
The error bars include statistical errors only. 
Curves (see text) correspond to 
the fits (B2)
(red), 
and (B4)
(blue), 
and
are extrapolated to the mass distributions edge points.
}
\label{fig:fit_miss}
\end{figure}
\begin{table}
\caption{PDG $P_{13}$(1720) parameters from fit (B) and new state parameters
from fit (C). Errors are statistical.\label{table:tabres_1}}
\vspace{2mm}
\begin{tabular}{|c|c|c|c|c|}  \hline
	          &  M (MeV)		 & $\Gamma$ (MeV)     & $\frac{\Gamma_{\pi \Delta}}{\Gamma}$ (\%) & $\frac{\Gamma_{\rho N}}{\Gamma}$ (\%)       \\ \hline
PDG $P_{13}$ (B)  &  1725$\pm$20	 & 114$\pm$19	      & 63$\pm$12			  	  & 19$\pm$9			 		\\ \hline
 PDG \cite{Pdg96}  &  1650-1750		 &     100-200	      &		N/A			  	  & 70-85		   			\\ \hline
new $P_{13}$ (C)  &  1720$\pm$20	 & 88$\pm$17	      & 41$\pm$13			  	  & 17$\pm$10			 		\\  \hline
\end{tabular}	
\end{table}

\begin{figure}[h]
\vspace{5.cm}                                                                  
\includegraphics{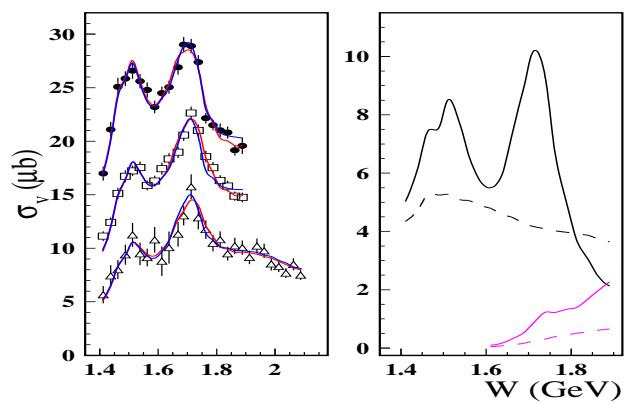}
\caption[]{Left: Total cross section for 
$\gamma_{v} p \rightarrow p \pi^{+} \pi^{-}$ as a function of $W$
from CLAS at the 3 mentioned $Q^{2}$ intervals (see fig.1). 
The error bars are statistical only.
The curves (see text) correspond to 
the fits 
(B2) (red),
and (B4) (blue).
Right: subdivision of the fitted cross section (B2) for
$Q^{2}$=0.5-0.8~(GeV/c)$^2$ into
resonant $\Delta^{++}\pi^{-}$ (black solid), continuum $\Delta^{++}\pi^{-}$
(black dashed), resonant $\rho^{0}p$ (magenta solid),
and continuum $\rho^{0}p$ (magenta dashed). Notice the different
vertical scales.
}
\label{fig:Xsec_allq2_data_and_bestnominal}
\end{figure}

To improve our
   reference curves before fitting the bump at around 1.7 GeV, the
   following steps were carried out:
the $P_{11}$(1440) strength was fitted to our low W data;
the $D_{15}$(1675) and the $D_{13}$(1700) photocouplings (which vanish in the
   SQTM) were replaced by NRQM values from Ref. \cite{Clo90}; 
an empirically established $A_{1/2,3/2}$ SQTM fitting uncertainty or
NRQM uncertainty of 10 or 20\% 
($\sigma$) was applied to all $N^{*}$ states; 
the hadronic parameters were allowed to vary for the $D_{13}$(1700) 
according to Ref. \cite{Man92}; and finally, the curves
providing the best $\chi^{2}/\nu$ were selected as the starting points.
Then, first we performed three fits, (B1), (B2), and (B3),
where the photocouplings of only one resonance at a time
were varied.
In (B1), we varied A$_{1/2}$, A$_{3/2}$, hadronic couplings,
and position of the D$_{13}$(1700) in a wide range.
In (B2), the same was done for the P$_{13}$(1720), 
and in (B3) for the P$_{11}$(1710).
In both fits (B2) and (B3),
we also varied the hadronic
parameters and the position of the D$_{13}$(1700)
over a range consistent with their large uncertainties
from Ref. \cite{Man92}. 
Fits (B1) and (B3) gave a poor description of the data,
with $\chi^2/\nu$=5.2 and 4.3, respectively.
The best fit ($\chi^2/\nu$ =3.4)
was obtained in (B2) (Fig.~\ref{fig:fit_miss}).
However, the resulting values for the branching fractions 
of the P$_{13}$(1720) were significantly
different from previous analyses
reported in the literature and well outside 
the reported errors \cite{Pdg96,Man92,Dyt00}. 
Starting from (B3), we then performed a final fit, 
(B4), for which the $P_{13}$(1720)
hadronic couplings were fixed from the literature, and  
varying the photocouplings of all three candidate states,
D$_{13}$(1700), P$_{13}$(1720), and $P_{11}$(1710), by 100\% ($\sigma$).
No better solution was found, 
the $\chi^2/\nu$ being 4.3 (Fig.~\ref{fig:fit_miss}).  
In Fig.~\ref{fig:Xsec_allq2_data_and_bestnominal} we report the final
comparison of fits (B2) and (B4) with the total cross section data.
Table~\ref{table:tabres_1} shows our results (first row) 
with statistical 
uncertainties, in comparison with the PDG values (second row). 
Our fits were not providing an unambigous
separation of $A_{1/2}$, $A_{3/2}$, 
and the longitudinal $S_{1/2}$, so we report as result
the total photocoupling strength, 
$\sqrt{A_{1/2}^{2} + A_{3/2}^{2}+ S_{1/2}^{2}}$.
Such value for the $P_{13}$(1720) fit is reported in the 
first three rows of
Table~\ref{table:tabres_2}.
The errors reflect the statistical uncertainties
   in the data and the correlations among the different resonances.

As discussed above, fitting the data around 1.7 GeV with 
   established baryon states leads either to a poor fit or to a drastic change
in resonance parameters with respect to published results.
In the framework of our analysis, there is
   no way to assess the reliability of the previously determined hadronic 
parameters of the PDG $P_{13}$(1720).
The resonant content of the reaction $\pi N \to \pi\pi N$,
                     which is used to obtain the hadronic parameters,
                     may be different from that of reactions initiated by
                     an electromagnetic probe. In particular, the $P_{13}$(1720) 
		     state seen in $\pi N \to \pi\pi N$
   may not be excited in electroproduction, while 
   some other state that decouples from $\pi N$ may be 
   excited electromagnetically.
	This possibility is studied in the next step.
%
\begin{table}
\caption{PDG $P_{13}$(1720) total photocoupling from fit (B2) 
and new state total photocoupling
from fit (C). Errors are statistical.\label{table:tabres_2}}
\vspace{2mm}
\begin{tabular}{|c|c|c|}  \hline
 step	   & $Q^{2}$	       &    $\sqrt{A_{1/2}^{2} + A_{3/2}^{2}+ S_{1/2}^{2}}$	   \\
	   &(GeV/c)$^{2}$      &($10^{-3}/\sqrt{{\rm GeV}}$)				   \\ \hline
 B2		   &	 0.65		   &	 83$\pm$5	     	        	      \\ \hline
 B2		   &	 0.95		   &	 63$\pm$8	     	        	      \\ \hline
 B2		   &	 1.30		   &	 45$\pm$27	     	        	      \\ \hline
 C		   &	 0.65		   &	 76$\pm$9	     	        	      \\ \hline
 C		   &	 0.95		   &	 54$\pm$7	     	        	      \\ \hline
 C		   &	 1.30		   &	 41$\pm$18	     	        	      \\ \hline
\end{tabular}		  
\end{table}

Step (C) - We investigated whether our
   data could be fitted by including another baryon state, 
while keeping the hadronic parameters of 
the $P_{13}$(1720) as in Refs. \cite{Pdg96,Man92}.
The quantum numbers $S_{I1}$,$P_{I1}$,$P_{I3}$,$D_{I3}$,$D_{I5}$,
$F_{I5}$,$F_{I7}$ were tested on an equal footing, where $I$/2
   is the isospin, 
   undetermined in our measurement.
We then simultaneously varied the photocouplings
   and the hadronic parameters of the new state and the $D_{13}$(1700).
   The total decay width of the new state was varied in the range of
   40-600 MeV, while its position was varied from 1.68-1.76 GeV. 
The best fit ($\chi^2/\nu$=3.3) was obtained with a $P_{I3}$ state, 
while keeping
the $P_{13}$(1720) hadronic parameters at published values.
Other partial waves gave a $\chi^{2}/\nu$ $\geq$ 4.2.
Curves obtained from the best fit were nearly identical with the 
red solid lines in
Figs.~\ref{fig:fit_miss} and~\ref{fig:Xsec_allq2_data_and_bestnominal}.
In order to avoid the unobserved $\rho$ production peak
(Fig.~\ref{fig:Xsec_allq2_data_and_nominal_and_pipi}, right), 
the photocouplings of the PDG $P_{13}$(1720)
had to be reduced by about a factor of two with respect to the SQTM prediction,
making its contribution very small.
Instead, in this fit the main contribution to the bump 
came from the new state.
Resonance parameters and total photocoupling value obtained for the hypothetical new state
are reported in Table~\ref{table:tabres_1} (last row) and~\ref{table:tabres_2} 
(last 3 rows), respectively.

A second $P_{13}$ state is indeed predicted in the quark model of 
Ref. \cite{Cap94}, however with a mass 
of 1870 GeV. 
The presence of a new three-quark state with the 
same quantum numbers as the conventional
$P_{13}$(1720) in the same mass range would likely lead to strong mixing. 
However, as mentioned above, a different
isospin and/or partial wave cannot be excluded. 
Finally, the new state may have a different internal structure, such as a 
hybrid baryon with excited glue components.
Such a $P_{13}$ hybrid state is predicted in the flux tube model \cite{Pag00}. 
Yet another possibility is that some resonance parameters established in 
previous analyses may have much larger uncertainties than reported in the 
literature.
In this case, outlined in our step (B),
our analysis would establish new, more precise parameters for a 
known state, and invalidate previous results.

In conclusion, in this letter we presented data on 
the $ep \to e'p \pi^+ \pi^-$ reaction in a wide kinematic range,
with higher quality than any previous double 
pion production experiment.
Our phenomenological calculations using existing PDG parameters 
provided a poor agreement with the new data at $W \sim$ 1700 MeV.  
We explored two alternative interpretations of 
the data. If we dismiss previously established hadronic 
parameters for the $P_{13}$(1720) we can fit the data with
a state having the same spin/parity/isospin but strongly 
different hadronic couplings from the PDG state. 
If, alternatively, we introduce a new state in addition 
to the PDG state with about the same mass, spin $\frac{3}{2}$, and 
positive parity, a good fit is obtained for a state having a 
rather narrow width, a strong $\Delta \pi$ coupling, and a 
small $\rho N$ coupling, while keeping
the PDG $P_{13}$(1720) hadronic parameters at published values. 
In either case we determined the total photocoupling 
at $Q^2 > 0$.
A simultaneous analysis of single and 
double-pion processes provides more constraints and may  
help discriminate better between alternative interpretations of 
the observed resonance structure in the CLAS data. Such 
an effort is currently underway.

We would like to acknowledge the outstanding efforts of the 
staff of the Accelerator
and the Physics Divisions at JLab that made this experiment possible. 
This work was supported in part by the Istituto Nazionale di Fisica Nucleare, 
the French Commissariat \`a l'Energie Atomique,  
the U.S. Department of Energy and National Science Foundation, 
and the Korea Science and Engineering Foundation.
U. Thoma acknowledges an ``Emmy Noether'' grant from the 
Deutsche Forschungsgemeinschaft.
The Southeastern Universities Research Association (SURA) operates the
Thomas Jefferson National Accelerator Facility for the United States
Department of Energy under contract DE-AC05-84ER40150.

\end{document}